# No-cloning principal can alone provide security

**Arindam Mitra**

Anushakti Abasan, Uttar-Phalguni-7, 1/AF, Salt Lake,
Kolkata, West Bengal, 700064, India.

Abstract: Existing quantum key distribution schemes need the support of classical authentication scheme to ensure security. This is a conceptual drawback of quantum cryptography. It is pointed out that quantum cryptosystem does not need any support of classical cryptosystem to ensure security. No-cloning principal can alone provide security in communication. Even no-cloning principle itself can help to authenticate each bit of information. It implies that quantum password need not to be a secret password. Ones name can be ones quantum password.

In classical cryptography, a key is needed to encrypt secret message. In 1949, Shannon proved [1] that one-time-pad [2] is an unbreakable cryptosystem provided key is not reused. Still, one-time-pad system is not popular because always new key is required.

Around 1970, Wisner first realized [3] that quantum no-cloning principle [4] could ensure security of data, although no-cloning principle was discovered later. In 1984, after the delayed publication of Wisner's seminal work [3], Bennett and Brassard discovered [5] a quantum key distribution (QKD) protocol where two distant users can generate secure key without secretly sharing any key. Since then many QKD protocols have been designed [6].

QKD system will completely break down [7], if eavesdropper impersonates legitimate receiver. To do so, eavesdropper has to intercept the transmitted quantum states. To foil this simple attack [7], Wegman-Carter classical authentication protocol [8], where classical laws of probability provide security, and which operates on a secretly shared key, can be incorporated within QKD system. It means QKD system quantum no-cloning principle and classical laws of probability jointly provide security. Of course this dependency on classical system will not be a problem for the implementation of QKD system. But, it would be always interesting to know whether quantum mechanics can alone ensure security in communication or whether no-cloning principle can be used to authenticate quantum communication.

In standard quantum coding a single quantum state is used to encode bit values. All the existing QKD systems are based on standard quantum coding. As an alternative, two sequences of quantum states can be used to represent two bit values. We shall see that arbitrarily long, secret and authenticated message can be transmitted repeatedly using two sequences of quantum states as described below.

1. Alice and Bob secretly share the preparation codes of the two sequences $S_0$ and $S_1$ representing 0 and 1 respectively.
2. Alice transmits either $S_0$ or $S_1$ to Bob.
3. Bob recovers the bit value encoded in the sequence.
4. Bob sends the complementary sequence to Alice, if he recovers the bit value without detecting any error. That is, if Bob gets $S_0$ he sends $S_1$ and vice versa. If Alice does not get any error, she sends the next bit to Bob.
5. If any of them detects error, they never use the same pair of sequences.

Note that each bit can be authenticated, if eavesdropper cannot break the sequence-code without introducing error. The step 4 ensures that eavesdropper cannot simultaneously possess more than one copy of any of the two sequences. It implies that eavesdropping is restricted by the no-cloning principle.

It is known that two non-orthogonal states can ensure security [9] of data since non-orthogonal states cannot be cloned. Therefore, it is obvious that a sequence (s) of quantum states, comprising with some non-orthogonal states, can also ensure security. It will be advantageous if we can utilize non-orthogonality as best as possible. Next we shall present such scheme.

Suppose Alice has a stock of horizontally polarized photons and two ideal coins 1 and 2. The photons can be transmitted through a symmetric (1:1) beam splitter leading to two paths "s" and "r". Alice prepares four superposition states in the following way. She tosses the coin 1 and if it yields "tail" (T) then in the path "s", she rotates the polarization by $90^0$ ($|\leftrightarrow\rangle_s \longrightarrow |\updownarrow\rangle_s$) or if "head" (H) she does nothing ($|\leftrightarrow\rangle_s \longrightarrow |\leftrightarrow\rangle_s$). In the path "r", she does nothing ($|\leftrightarrow\rangle_r \longrightarrow |\leftrightarrow\rangle_r$). Simultaneously, by a phase shifter a phase factor + 1 or -1 is introduced at random in the path "s" tossing coin 2. If the tossing of coin 2 yields "head", phase factor + 1 or if "tail" phase factor - 1 is introduced. After the double tossing she allows photon to proceed through the beam splitter. The double outcomes of double tossing can be denoted as $H_H$, $H_T$, $T_H$ and $T_T$. The four superposition states prepared in this way may be written as

$$H_H \longrightarrow |A_+\rangle_i = \frac{1}{\sqrt{2}}(|\leftrightarrow\rangle_r + |\leftrightarrow\rangle_s)$$

$$H_T \longrightarrow |A_-\rangle_i = \frac{1}{\sqrt{2}}(|\leftrightarrow\rangle_r - |\leftrightarrow\rangle_s)$$

$$T_H \longrightarrow |B_+\rangle_i = \frac{1}{\sqrt{2}}(|\leftrightarrow\rangle_r + |\updownarrow\rangle_s)$$

$$T_T \longrightarrow |B_-\rangle_i = \frac{1}{\sqrt{2}}(|\leftrightarrow\rangle_r - |\updownarrow\rangle_s)$$

Alice prepares a sequence $S_0$ of the above states. To prepare $S_0$, she has to toss each coin N times. To prepare another sequence of another set of superposition states at random, she proceeds in the following way. If the tossing of coin 1 yields "head" (H) then in the path "s", she rotates the polarization by $45^0$ ($|\leftrightarrow\rangle_s \longrightarrow |\nearrow\rangle_s$) or if "tail" (T), rotates by $135^0$ ($|\leftrightarrow\rangle_s \longrightarrow |\nwarrow\rangle_s$). Similarly, in the other path "r", she does nothing. Simultaneously, she introduces a phase factor + 1 or -1 at random in the path "s" tossing coin 2. The four states prepared in this way may be written as

$$H_H \longrightarrow |C_+\rangle_i = \frac{1}{\sqrt{2}}(|\leftrightarrow\rangle_r + |\nearrow\rangle_s)$$

$$H_T \longrightarrow |C_-\rangle_i = \frac{1}{\sqrt{2}}(|\leftrightarrow\rangle_r - |\nearrow\rangle_s)$$

$$T_H \longrightarrow |D_+\rangle_i = \frac{1}{\sqrt{2}}(|\leftrightarrow\rangle_r + |\nwarrow\rangle_s)$$

$$T_T \longrightarrow |D_-\rangle_i = \frac{1}{\sqrt{2}}(|\leftrightarrow\rangle_r - |\nwarrow\rangle_s)$$

Alice prepares a sequence, $S_1$, of the above states. To prepare $S_1$, she has to toss each coin N times. Note that, among the eight superposition states $|A_+\rangle$ and $|A_-\rangle$ are orthogonal to each other and $|B_+\rangle$ and $|B_-\rangle$ are orthogonal to each other and so on, but $|A_\pm\rangle, |B_\pm\rangle, |C_\pm\rangle, |D_\pm\rangle$ are mutually non-orthogonal to each other.

Alice transmits either $S_0$ or $S_1$ through the path "s" and "r" leading to Bob's site. Suppose Bob sets an analyzer either at $0^0$ or at $45^0$ at the end of the path "s". The projection operators corresponding to these two settings are denoted by $P_s^A$ and $P_s^C$. If $P_s^A$ and $P_s^C$ are applied on $|D_\pm\rangle_i$ and $|B_\pm\rangle_i$ respectively, the measurement will yield positive result with *a priori* probability 1/4. If $P_s^A$ and $P_s^C$ are applied on $|B_\pm\rangle_i$ and $|D_\pm\rangle_i$ respectively, the measurement will yield null result with *a priori* probability 1.

Bob can be able to identify the two sequences as he knows the sequence codes. Suppose he applies $P_s^A$ and $P_s^C$ at random in those events where all the states will be either $|B_\pm\rangle_i$ or $|D_\pm\rangle_i$. Only one of the two operators will yield positive result with probability 1/8. However, a single positive result is enough to identify the sequence. Suppose $P_s^A$ gives positive results. Bob correctly identifies the states as $|D_\pm\rangle_i$ and the sequence as $S_1$.

Suppose Alice transmits a single bit. Eavesdropper may try to recover the bit value encoded in the transmitted sequence. As eavesdropper does not know the sequence codes, all random sequences are equally probable to him. To eavesdropper, $S_0$ is a random *mixture* of states $|A_\pm\rangle_i$ and $|B_\pm\rangle_i$, and $S_1$ is also a random *mixture* of states $|C_\pm\rangle_i$ and $|D_\pm\rangle_i$. To eavesdropper, the density matrices of these two *mixtures* in the basis $\{|\leftrightarrow\rangle_r, |\updownarrow\rangle_r, |\leftrightarrow\rangle_s, |\updownarrow\rangle_s\}$ are same.

$$\rho_0^E = \rho_1^E = \frac{1}{4}\begin{pmatrix} 2 & 0 & 0 & 0 \\ 0 & 0 & 0 & 0 \\ 0 & 0 & 1 & 0 \\ 0 & 0 & 0 & 1 \end{pmatrix}$$

The following observations can be made.

1. The sequences $S_0$ and $S_1$ cannot be distinguished [10] by eavesdropper since $\rho_0^E = \rho_1^E$. Eavesdropper cannot recover bit value from the single copy of any transmitted sequence. Eavesdropper cannot gain any information of the sequence codes, if he disturbs only a single sequence.

2. Due to the restrictions, stated in the step 4, eavesdropper is bound to introduce error with non-zero probability, if he disturbs a sequence or a state since in each sequence out of the four states two states are non-orthogonal to each other, and the states of the one sequence are non-orthogonal to the states of the other sequence.

3. In the classical world, we know that whenever an adversary successfully impersonates he gains full information. This is true even for raw QKD protocols. But, in our system eavesdropper can successfully impersonate without gaining any information. Suppose Bob transmits the same sequence which he received. After intercepting the Alice's sequence eavesdropper can send it back to Alice without disturbing it. Alice will think that Bob got her bit and continue the transmission, if Bob cannot communicate with her through an authenticated channel. Therefore, eavesdropper can successfully impersonate as Bob to Alice without gaining any information.

In step 4, Bob sends complimentary sequence to Alice. If Alice recovers the complimentary sequence without detecting any error, she will be sure that Bob received her transmitted sequence. But Alice cannot always transmit complimentary sequence. Due to this reason, eavesdropper has the opportunity of sending back Bob's sequence to Bob without disturbing it and thereby masquerades as Alice. But this is possible for a single bit, not for more than one bit. So, it is not possible to impersonate as Alice before Bob and as Bob before Alice, if more than one bit is transmitted.

4. Given an unknown sequence comprising non-orthogonal quantum states, it cannot be cloned. But, here eavesdropper can gain information because the same sequence is used more than one time. It implies that two sequences of quantum states may not be described as random *mixture*, if they are used more than one time. Be that as it may, more the same sequence will be used, more the eavesdropper's information gain will increase. Of course more eavesdropper will try to gain information, more he will introduce error. Large probability of identification of states implies that large error will be introduced if the states are disturbed.

It is easy to see that eavesdropper will introduce error with probability ($p_e$) 1/8, if he intercepts a sequence of states and resends another random sequence of the same states. If this intercept-resend strategy is pursued, Bob's both the operators, $P_s^A$ and $P_s^C$, if used at random, will give positive result with probability 1/8. That is, $p_e = 1/8$.

5. The step 4 ensures that eavesdropper cannot break the sequence codes without introducing error. There is a caveat. For any quantum cryptosystem, eavesdropper's probability of introducing zero-error is $p_{ne} = (1 - p_e)^N$ when N states are disturbed. By increasing N this probability can be made arbitrarily small. For the presented protocol, if eavesdropper pursues the intercept-resend strategy as described above $p_{ne} \approx 10^{-6}$ for N = 100. So, if no error is detected the received sequence/bit is authenticated with probability nearly 1. Essentially, bit-by-bit security can be achieved through bit-by-bit authentication. A bit of information can be simultaneously made secure and authenticated. Since each bit can be made authenticated quantum password need not to be a secret password since each bit can be made authenticated. Ones name can be ones quantum password.

6. Quantum cryptography can evade Shannon's restriction [1] regarding the reuse of the same secret key(s).

In presence of noise, eavesdropper has the opportunity of storing some of the states of each sequence. Therefore, he gets more than one copy of some of the states. So, in noisy environment the advantage of no-cloning principle is somewhat lost even under bit-by-bit security criterion. It has been observed that entanglement-based alternative quantum coding (AQC) is possible [11]. To tackle noise, entanglement purification technique [12] might be incorporated within our entanglement-based AQC.

Will no-cloning principle itself be able to provide complete security in presence of noise ? The answer is yet to be known.

Acknowledgement: I am indebted to S. Wisner for voluntarily offering some financial help to support this work.

Email:mitra1in@yahoo.com